\documentclass[prl,showpacs,twocolumn]{revtex4-1}
\usepackage{graphicx,amssymb,color}
\usepackage[dvipsnames]{xcolor}

\begin{document}
\title{Fluctuations and Criticality of a Granular Solid-Liquid-like Phase Transition}
\author{Gustavo Castillo}
\author{Nicol\'as Mujica}
\email[Corresponding author: ]{nmujica@dfi.uchile.cl} 
\author{Rodrigo Soto}
\affiliation{Departamento de F\'isica, Facultad de Ciencias F\'isicas y 
Matem\'aticas, Universidad de Chile, Avenida Blanco Encalada 2008, Santiago, Chile}
\date{\today}

\begin{abstract}

We present an experimental study of density and order fluctuations in the vicinity of the solid-liquid-like transition that occurs in a vibrated quasi-two-dimensional granular system. The two-dimensional projected static and dynamic correlation functions are studied. We show that density fluctuations, characterized through the structure factor, increase in size and intensity as the transition is approached, but they do not change significantly at the transition itself. The dense, metastable clusters, which present square symmetry, also increase their local order in the vicinity of the transition. This is characterized through the bond-orientational order parameter $Q_4$, which in Fourier space obeys an Ornstein-Zernike behavior. Depending on filling density and vertical height, the transition can be of first or second order type. In the latter case, the associated correlation length $\xi_4$, relaxation time $\tau_4$, zero $k$ limit of $Q_4$ fluctuations (static susceptibility), the pair correlation function of $Q_4$, and the amplitude of the order parameter obey critical power laws, with saturations due to finite size effects. Their respective critical exponents are $\nu_{\bot} = 1$, $\nu_{||} = 2$, $\gamma = 1$, $\eta=0.67$, and $\beta=1/2$, whereas the dynamical critical exponent $z = \nu_{||}/\nu_{\bot}  = 2$. These results are consistent with model C of dynamical critical phenomena, valid for a non-conserved critical order parameter (bond-orientation  order) coupled to a conserved field (density).

\end{abstract}
\pacs{
64.60.Ht,
45.70.-n,
05.40.-a,
64.70.qj	
}
\maketitle

\paragraph{Introduction.}

A dry granular system is an a-thermal collection of macroscopic particles that interact mainly through dissipative, hard core-like, collisions. They behave as solids, liquids or gases depending on the nature of the forces that act upon them and the energy injection rate~\cite{RMP96}. These systems present phase transitions and coexistence. Simple examples are: a thin layer of vibrated sand, which for small driving amplitudes remains solid but can be completely fluidized for larger accelerations; an avalanche driven by gravity were a thin layer of grains flows above an almost solid pile. Granular systems are therefore excellent candidates for studying non-equilibrium phase transitions \cite{lubeck,Takeuchi}. In vibrated thin layers energy is transferred from the top and bottom lids to the vertical motion of the grains, which later transfer the energy to the horizontal motion at collisions, that are also dissipative. The sequence breaks detailed balance keeping the system out of equilibrium.

Recently, several granular systems that undergo interesting phase transitions have been reported \cite{olafsen,Argentina2002,transinstabilityMeerson,transinstabilityBrey,electrostatically,cartes,prevost2004,Melby2005,Olafsen2005,Reis2006,clerc2008,tanaka2008}. One particular system is a vibrated fluidized granular monolayer composed of $N$ hard spheres of diameter $d$ confined in a shallow cell of height $L_z<2d$ (typically $L_z\approx 1.7d-1.9d$). Under proper conditions, solid and liquid phases can coexist at mechanical equilibrium \cite{olafsen,prevost2004,Melby2005,clerc2008}. The solid clusters can present different order symmetries, like square or hexagonal, depending on forcing, geometrical and particle parameters. It has been reported that for $L_z \approx 1.7d - 1.8d$ and for a large range of filling densities, the most compact structure in quasi-2D is made of two layers of square symmetry. The more compact hexagonal structure formed by two layers needs a larger vertical gap or larger densities \cite{Melby2005}. The critical amplitude above which there is coexistence decreases with increasing density. 

Many of the previous works on granular phase transitions focus on the similarities or comparisons of such non-equilibrium systems with equilibrium phase transitions \cite{olafsen,prevost2004,Melby2005,Olafsen2005,Reis2006,tanaka2008}. 
For example, the equilibrium KTHNY theory has proved useful in the two dimensional melting of granular monolayers~\cite{Olafsen2005}.
Here we focus on a dynamical critical phase transition in a non-equilibrium quasi-2D granular system. We present an experimental study of the solid-liquid phase transition in a vibrated fluidized granular monolayer. The solid phase consists on two square interlaced layers, stabilized by the collisions with the top and bottom walls and the confining pressure exerted by the liquid phase~\cite{Melby2005,clerc2008}.
We focus on  density and bond-orientation order fluctuations in the vicinity of the transition. We show that the transition can be continuous or abrupt depending on the cell's height and filling density. Density fluctuations show a crossover behavior at the transition, whereas order shows strong fluctuations. In the continuous case several magnitudes show critical-like behavior being possible to measure five independent critical exponents. These results are consistent with model C of dynamical critical phenomena~\cite{hohenberg}, valid for a non-conserved critical order parameter (bond-orientation  order) coupled to a conserved field (density).

\begin{figure*}[t!]
\begin{center}
\includegraphics[width=.68\columnwidth]{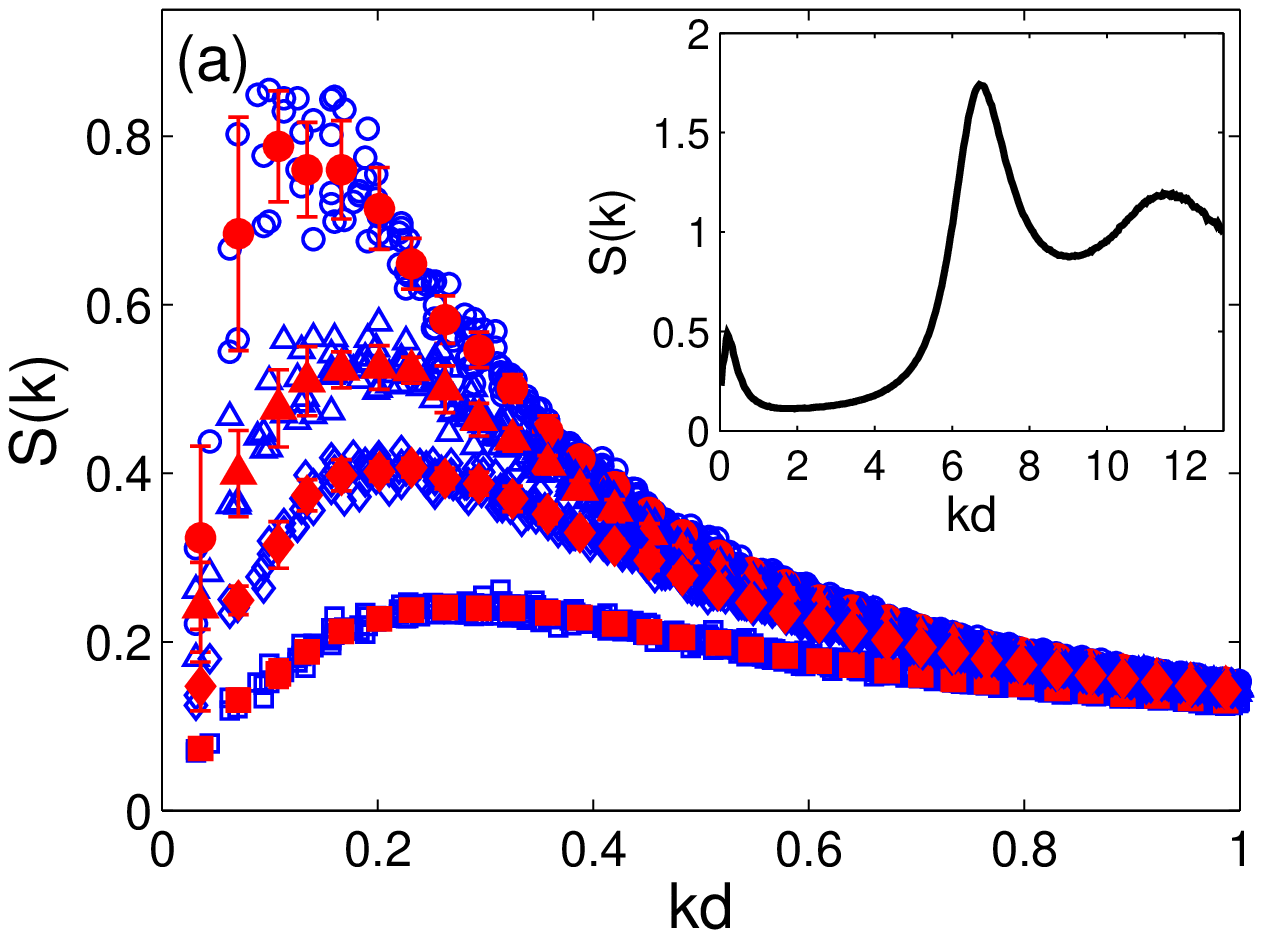}
\includegraphics[width=.68\columnwidth]{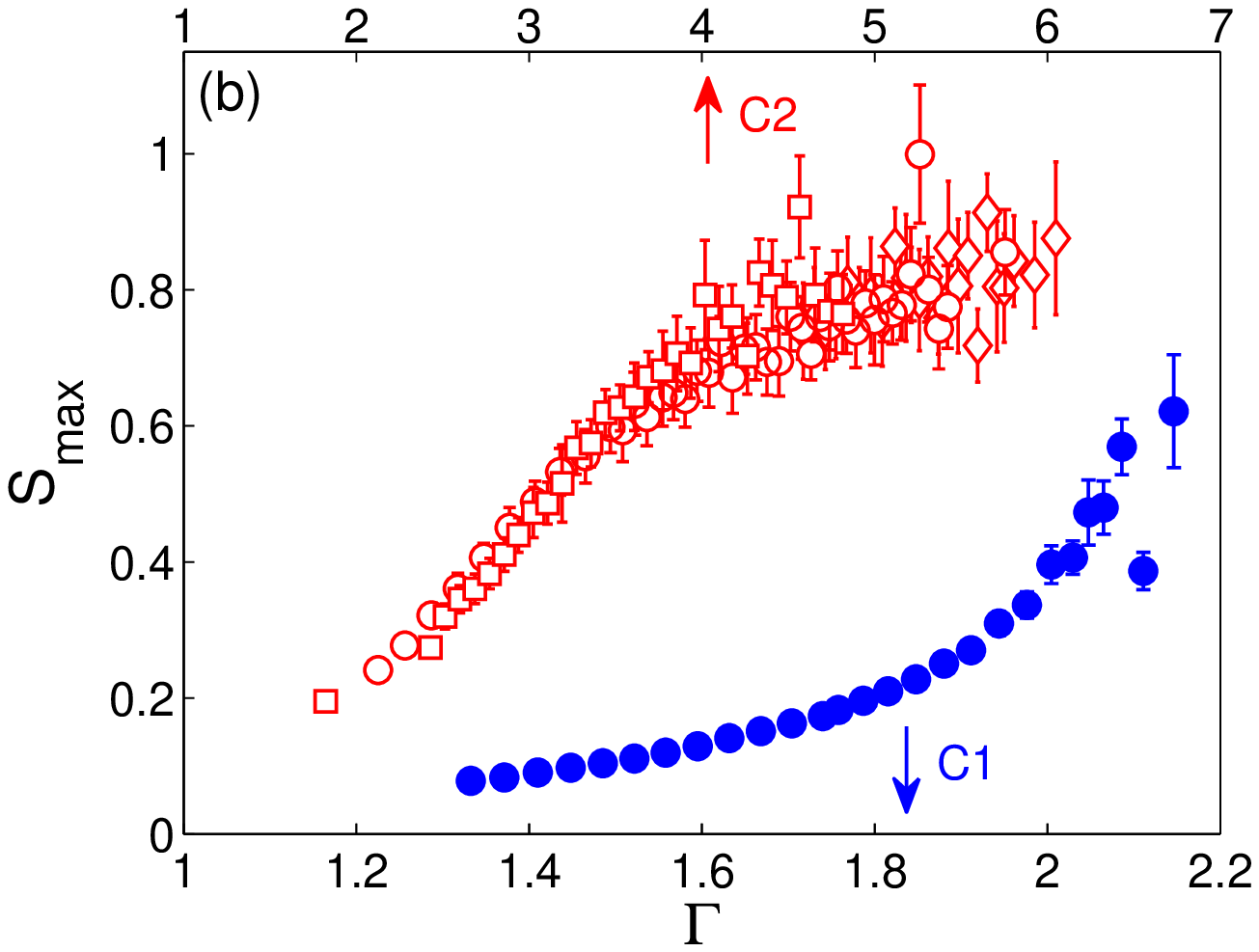}
\includegraphics[width=.68\columnwidth]{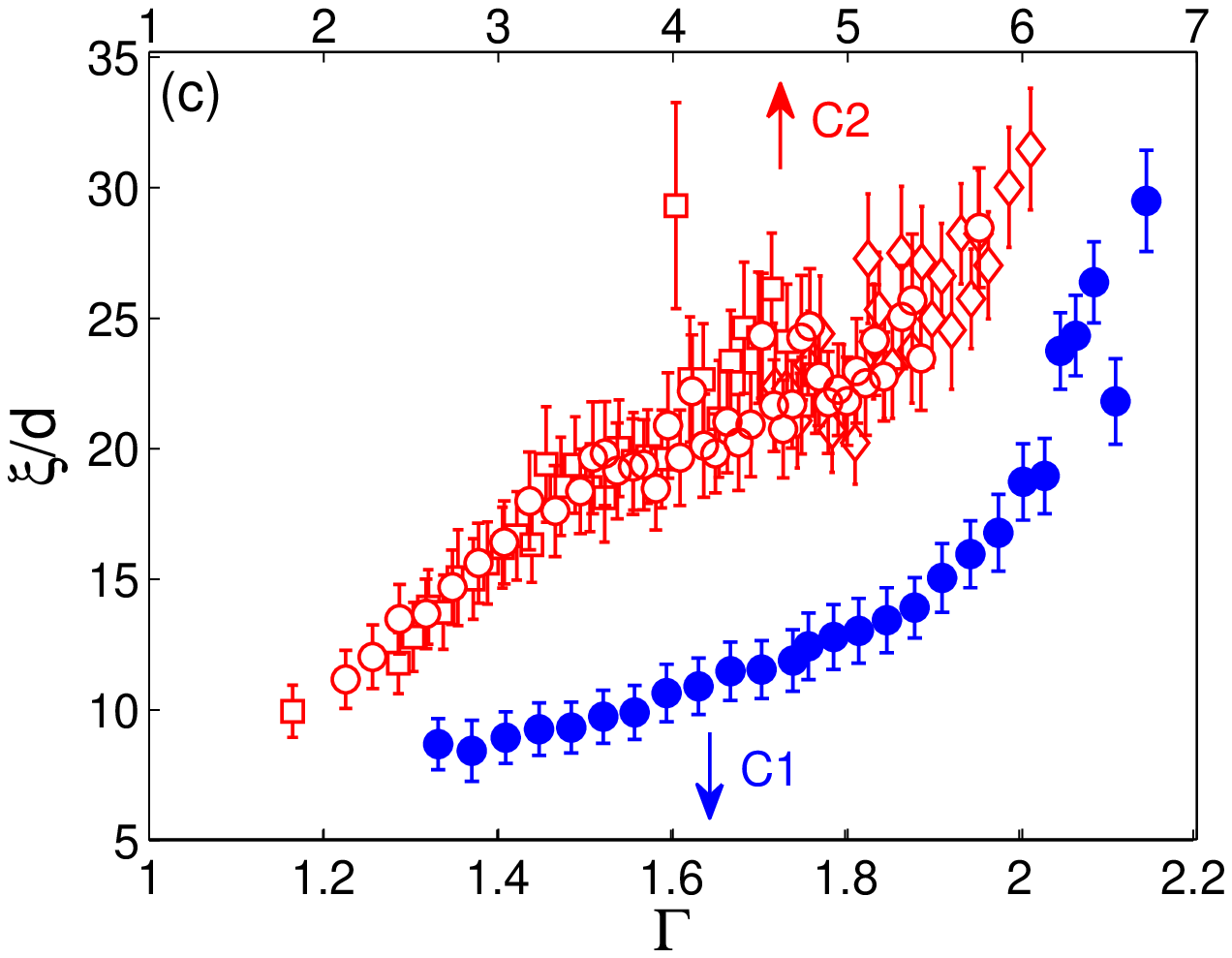}
\caption{{(color online). (a) $S(k)$ in the large wavelength limit for four different accelerations, $\Gamma = 2.13$, $2.74$, $3.18$ and $4.79$ (C2). Open symbols correspond to raw $S(k)$ data, whereas solid symbols correspond to averages using windows $k \in [n k_{\rm min},(n+1) k_{\rm min}]$ for integer $n\geqslant 1$, where $k_{\rm min} =  \pi / L$. Error bars correspond to standard deviations. The inset shows $S(k)$ for a larger range of $k$  for $\Gamma = 3.64$. (b) Pre-peak maximum $S_{\rm max} \equiv S(k=k^*)$, which occurs at $k = k^*$, and (c) associated length scale $\xi = \pi/k^*$ as functions of $\Gamma$ for C1 (open symbols) and C2 (closed symbols).}} 
\label{fig1}
\end{center}
\end{figure*}

\paragraph{Experimental setup and procedures.}

The experimental device is similar to the one described in \cite{rivas}. The granular system is composed by $N \sim 10^4$ stainless steel spherical particles, of diameter $d=1$ mm. The quasi-two-dimensional box has lateral dimensions $L_x = L_y \equiv L = 100d$.  In order to study the different possible transitions, first or second order type, two configurations are used: $L_z = 1.83d \pm 0.02d$ and $N = 9878$, or   $L_z = 1.94d \pm 0.02d$ and $N = 11504$. These are labeled configurations 1 (C1) and 2 (C2) respectively. The surface coverage is defined by the filling fraction $\phi = N\pi d^2/4L^2$ ($\phi_{\rm C1} = 0.776$ and $\phi_{\rm C2} = 0.904$).

The whole setup is forced sinusoidally with an electromechanical shaker, with displacement $z(t) = A \sin(\omega t)$. Top view images are obtained with a camera at $10$ fps. Particle positions are determined at
sub-pixel accuracy.  Results have been obtained by fixing the particle number $N$, cell height $L_z$ and driving frequency $f=\omega/2\pi = 1/T=80$ Hz. 
The dimensionless acceleration $\Gamma = A\omega^2/g$ is varied in the range $1-6$. 

\paragraph{Static structure function.}
Particle positions $\vec r_j(t)$ in the plane $(x,y)$ are determined for each time $t$.
Experimentally, there is no access to the $z$ coordinate. Thus, the 2D microscopic density field Fourier components are 
\begin{eqnarray}
\widehat{\rho}(\vec k,t) &=& \int d^2r\, e^{i\vec r\cdot\vec k} \rho(\vec r,t)=\sum_{j=1}^{N} e^{i \vec k \cdot \vec r_j(t)}.
\end{eqnarray}
The static structure factor $S(\vec k)$ measures the intensity of density fluctuations in Fourier space:
\begin{eqnarray}
S(\vec k) &=& \frac{\langle|\widehat\rho(\vec k,t) - \langle \widehat\rho(\vec k,t) \rangle|^2\rangle}{N}, \label{defSk}
\end{eqnarray}
where $\langle \,\, \rangle$ denotes time averaging. In general $\langle \rho(\vec k) \rangle \neq 0$, due to inhomogeneities induced by boundary conditions. The wave vectors are computed from $\vec k = \pi ( n_x \hat {\i} +  n_y\hat  {\j})/L$, where $n_x, n_y \in \mathbb{N}$.

In the liquid phase ($\Gamma < \Gamma_c$) we have verified that the system is isotropic, $S(\vec k) = S(k)$, where $|\vec k| \equiv k$. In the phase separated regime this quantity is not well defined as density fluctuations should in principle behave differently in each phase. However, as it is not easy to determine $\Gamma_c$ precisely, we use this quantity as a possible relevant measurement in the vicinity of the solid-liquid phase separation, even above the critical amplitude. 

Figure \ref{fig1}a presents $S(k)$ obtained for C2 (the qualitative features are the same for both configurations and their differences will pointed out explicitly later). The main figure presents the long-wavelength range, $kd \leqslant 1$, for four $\Gamma$ below $\Gamma_c$. The inset presents $S(k)$ for a larger range of $k$. It has the usual form expected for liquids with short range order, but with a pre-peak located in the range $kd = 0.1 - 0.3$. The associated density fluctuations  are indeed visible by simple visual inspection~\cite{suppmat}. 

The pre-peak is characterized by its maximum value at $k^*$, $S_{\rm max} \equiv S(k^*)$, and the associated characteristic length scale $\xi = \pi/k^*$. These quantities are plotted in Fig. \ref{fig1}b-c as functions of $\Gamma$ for increasing amplitude ramps and for both configurations. They both increase as the transition is approached. The difference between configurations is mainly manifested in the shape of each curve, being their final values (near the transition) very similar, $S_{\rm max} \approx 0.5 -0.8$ and $\xi/d \approx 20-30$. By observing visually the persistence  of the solid clusters we conclude that for C1 the transition is located at $\Gamma_c \approx 2$. For C2 it is more difficult to determine with the same precision but it is found to be $\Gamma_c \sim 5$. However, neither $S_{\rm max}$ or $\xi$ show evident changes at these values.

Density fluctuations do not show critical behavior, but they are needed to create regions of high order. Similar density fluctuations have been observed in amorphous materials \cite{Elliott1991,Tanaka2005b}, which have been consistently related to the existence of medium-range-crystalline-order. In our case, medium range order will be analyzed with an appropriate bond-orientational order parameter, which presents critical behavior. 

\paragraph{Bond-orientational order parameter.}

\begin{figure}[t!]
\begin{center}
\includegraphics[width=\columnwidth]{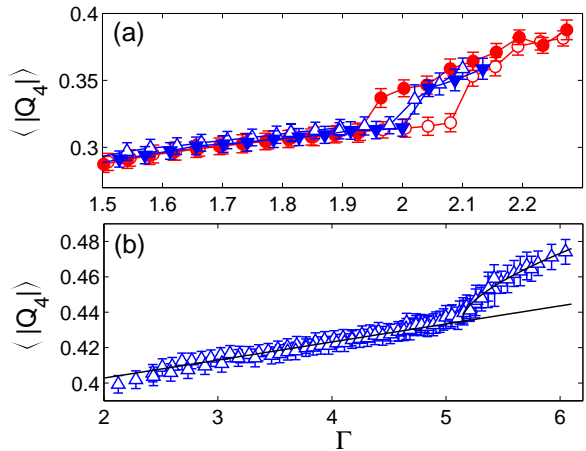}
\caption{(color online). Average global 4-fold bond-orientational order parameter $\langle |Q_4| \rangle$ versus  $\Gamma$ for C1 (a) and C2 (b). Open (solid) symbols represent data obtained for increasing (decreasing) $\Gamma$ ramps, with the following rates: $\Delta\Gamma/\Delta t \approx 0.005$ min$^{-1}$ ({\color{blue} $\triangle$, $\blacktriangledown$}) and $\Delta\Gamma/\Delta t \approx 0.02$ min$^{-1}$ ({\color{red} $\circ$, $\bullet$}). Continuous lines in (b) correspond to fits of the linear trend $Q_4^L = a\Gamma +b$ for $2.5<\Gamma<5$, with $a=0.011\pm0.001$ and $b=0.380\pm0.002$, and a supercritical-like behavior $ \langle |Q_4| \rangle=Q_4^L + c(\Gamma-\Gamma_c)^\beta$, with $\beta=1/2$, observed for $\Gamma \gtrsim 5$. }
\label{fig2}
\end{center}
\end{figure}

In the vicinity of the transition, fluctuations of high density present the same square symmetry as the solid phase. In the quasi-2D geometry the solid phase consists of two square interlaced layers instead of the hexagonal layer that is characteristic of 2D systems~\cite{Melby2005}. The local order can be characterized through a 4-fold bond-orientational order parameter. This is still valid in quasi-2D geometry because the interlaced two-layer square lattices (with unit cell length $d$ in each plane) result also in a square lattice when projected in 2D, with unit cell length $\sqrt{2}d/2$ when the grains are close packed. The 4-fold bond-orientational order parameter per particle is defined~\cite{resQ6}
\begin{equation}
Q_4^j = \frac{1}{N_{j}} \sum_{s = 1}^{N_{j}} e^{4i\alpha_{s}^j},
\end{equation}
where $N_j$ is the number of nearest neighbors of particle $j$ and $\alpha_s^j$ is the angle between the neighbor $s$ of particle $j$ and the $x$ axis.  For a particle in a square lattice, $|Q_4^j | = 1$ and the complex phase measures  the square lattice orientation respect to the $x$ axis. The corresponding global average and Fourier components are 
\begin{equation}
\langle |Q_4| \rangle = \left \langle \frac{1}{N} \sum \limits_{j=1}^{N} |Q_4^j| \right \rangle, \,
\widehat{Q}_4(\vec k,t) = \sum_{j=1}^{N} Q_4^j  e^{i \vec k \cdot \vec r_j(t)}.
\end{equation}
The average $\langle |Q_4| \rangle $  measures the fraction of particles in the ordered phase. This quantity is presented in Fig. \ref{fig2} as function of $\Gamma$ for both configurations. Results for increasing (decreasing) $\Gamma$ ramps are represented by open (solid) symbols. Two ramp rates are also reported: {\it slow ramps}, for which a quasi-static state has been reached, and {\it fast ramps}, for which it has not. The difference between both configurations is evidenced in the jump of about $10\%$ that is measured for $\langle |Q_4| \rangle $ at the transition for C1. Moreover, the position of this jump depends on the $\Gamma$ ramp rate: for the slow rate the increasing and decreasing ramp jumps coincide, whereas for faster ramps the increasing (decreasing) ramp jump occurs at higher (lower) $\Gamma$. We use the slow ramp data to obtain a measurement of the critical acceleration, $\Gamma_c^{\rm C1} = 2.01\pm 0.03$. By the contrary, the results obtained for C2 show first a linear trend for low $\Gamma$ and a clear deviation around $\Gamma \approx 5.1$, with no measurable jump. In fact, the deviation from the linear behavior obeys a supercritical-like law. For $\Gamma>5.2$ we have fitted the data with the function $\Delta Q_4 = \langle |Q_4| \rangle  - Q_4^L =  c (\Gamma-\Gamma_c)^\beta$, where $Q_4^L$ is the extrapolation of the linear trend observed for lower $\Gamma$. We obtain $c=0.029\pm0.002$, $\Gamma_c^{\rm C2} = 5.12\pm0.01$, and the exponent of the order parameter amplitude is $\beta=1/2$. Within experimental errors, the decreasing ramps also coincide with the increasing ramps in this configuration \cite{suppmat}. Consequently, the transition for configuration C1 is abrupt, of first-order type, whereas for C2 it is continuous, of second-order type. 

Local order can also be analyzed through its fluctuations in Fourier space by means of the 4-fold bond-orientational structure factor
\begin{equation}
S_4(\vec k) = \frac{\langle |\widehat Q_4(\vec k,t) - \langle \widehat Q_4(\vec k,t) \rangle|^2\rangle}{N}. \label{defSk4}
\end{equation}

\begin{figure}[ht!]
\begin{center}
\includegraphics[width=\columnwidth]{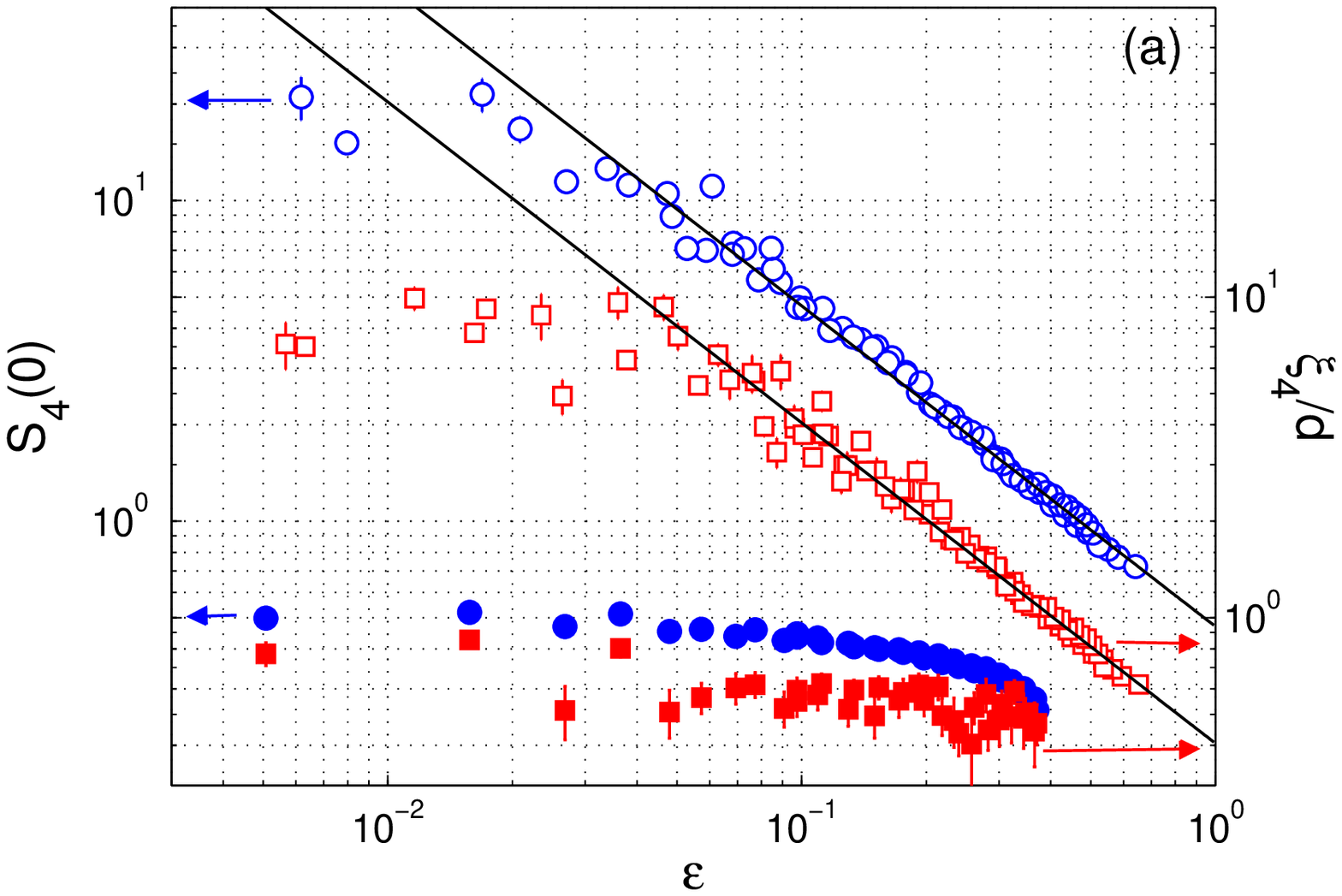}
\includegraphics[width=\columnwidth]{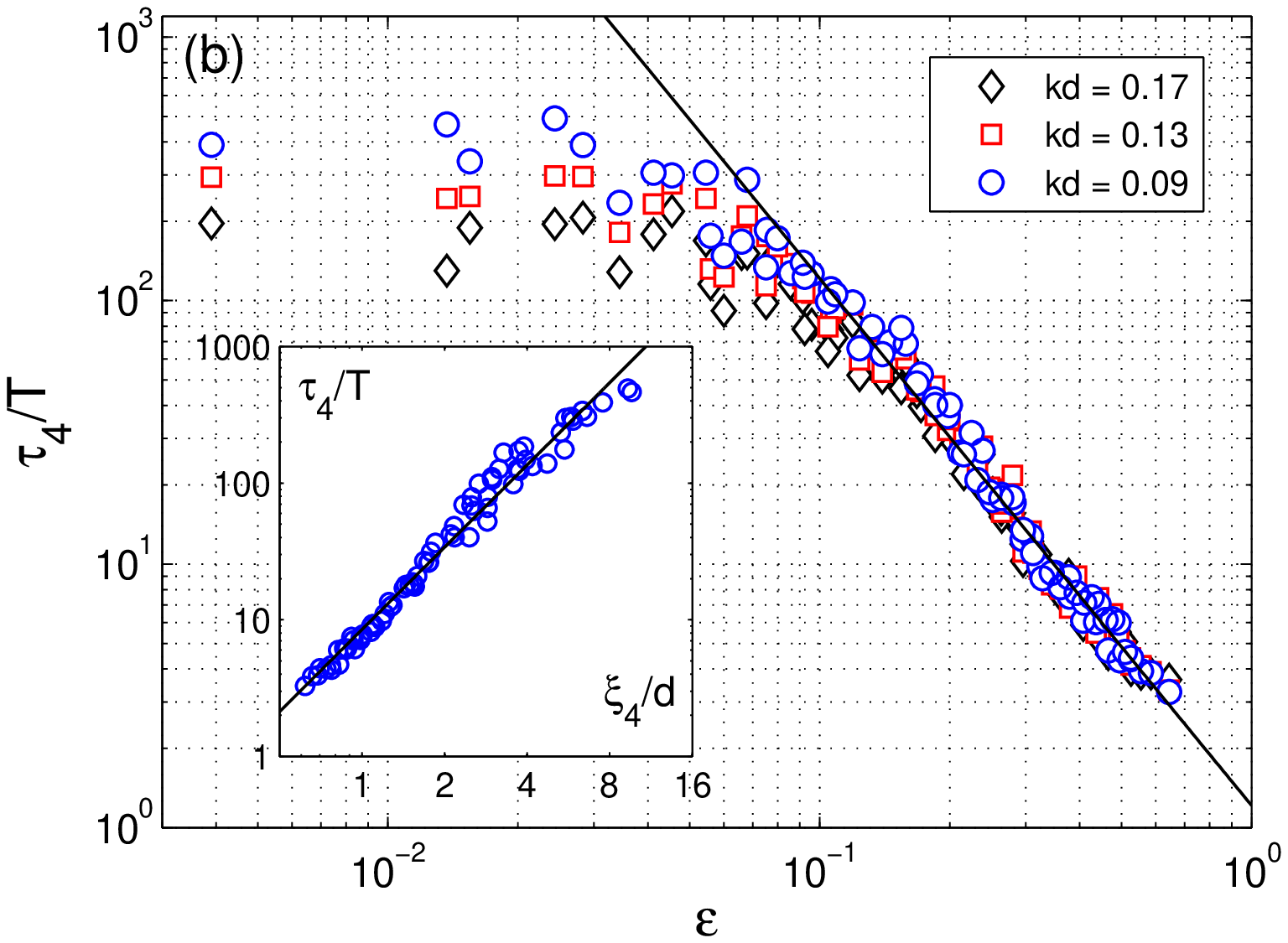}
\caption{(color online). (a) $S_4(0)$ ($\color{blue} \circ, \bullet$) and $\xi_4/d$  ($\color{red} \square, \blacksquare$) versus $\varepsilon$ for C1 (solid symbols) and C2 (open symbols). (b) $\tau_4/T$ versus $\varepsilon$ for C2, for three different low $k$. Continuous lines show critical power law fits, with exponents $\gamma = \nu_{\bot} = 1$ for $S_4(0)$ and $\xi_4$, and $\nu_{||} = 2$ for $\tau_4$. The fitted critical accelerations are $\Gamma_c^{\rm C2} = 5.09\pm0.07$, $\Gamma_c^{\rm C2} = 5.24\pm0.08$ and $\Gamma_c^{\rm C2} = 5.12\pm0.07$ respectively. The inset of (b) presents $\tau_4$ versus $\xi_4$ for $\varepsilon > 0.03$. The continuous line shows a fit  $\tau_4 \sim (\xi_4)^z$, with the dynamical exponent $z = \nu_{||}/\nu_{\bot} = 2$. }
\label{fig3}
\end{center}
\end{figure}

\noindent
For both configurations and for $\Gamma<\Gamma_c$, $S_4(k)$ shows an Ornstein-Zernike-like behavior in the limit $kd\ll1$ \cite{suppmat},
$S_4(k) \approx S_4(0)/[1+(\xi_4 k)^2]$,
where $\xi_4$ and $S_4(0)$ are the 4-fold bond-orientational correlation length and static susceptibility respectively. 

For configuration C1, the 4-fold bond-orientational susceptibility $S_4(0)$ and normalized correlation length $\xi_4/d$ vary weakly as the transition is approached (Fig. \ref{fig3}a). Defining the reduced acceleration $\varepsilon = (\Gamma_c-\Gamma)/\Gamma_c$, we obtain that for $0.005<\varepsilon<0.4$, $S_4(0)$ and $\xi_4/d$ vary in the ranges $0.25-0.5$ and $0.5-0.7$ respectively.  The fact that $S_4(0)<1$ and $\xi_4/d<1$ implies that fluctuations of the global 4-fold bond-orientational order parameter are weak and that there is practically no order correlation below the first-order-type transition. 

For C2 the situation is markedly different. Fig. \ref{fig3}a shows that $S_4(0)$ and $\xi_4/d$ vary strongly as the transition is approached. In the limit $\varepsilon \rightarrow 0$ they both saturate, presumably due to the system's finite size. For $\varepsilon \lesssim 3\times10^{-2}$ they saturate to $S_4(0)\approx 20$ and $\xi_4/d\approx 10$ respectively. This figure also demonstrates that both quantities follow the critical-like behavior,
\begin{equation}
S_4(0) = \tilde a \varepsilon^{-\gamma},\quad\quad\quad
\xi_4/d = \tilde b \varepsilon^{-\nu_{\bot}},
\end{equation}
with the critical exponents $\gamma=1$ and $\nu_{\bot}=1$.
The critical divergence with $\varepsilon$ makes it necessary to fit the $\Gamma_c$ separately for each case; for details on the procedure see the Supplemental Material \cite{suppmat}. The adjusted critical accelerations are $\Gamma_c^{\rm C2}=5.09~\pm~0.07$ and $\Gamma_c^{\rm C2} = 5.24\pm0.08$ respectively. Within experimental errors both critical accelerations are very consistent, as well as with the value obtained from the supercritical behavior of $\Delta Q_4$ ($\Gamma_c^{\rm C2} = 5.12\pm0.01$). Notice that now these are obtained from fits of measured quantities below the transition, whereas before, it was obtained with a fit of the order parameter above the transition. 

In the hydrodynamic regime ($d/\xi_4\lesssim k d\ll 1$) and in the vicinity of the transition, the bond-orientational structure factor is theoretically expected to behave as $S_4(k)=C_{\infty} k^{-(2-\eta)}$. When $\Gamma=5.10$ a power-law behavior is indeed observed in the range $0.1\leq kd\leq 1$ with an anomalous exponent 
$\eta=0.67\pm0.01$~\cite{suppmat}.

As a final evidence of the observed criticality we now turn to the characterization of the relaxation time of the metastable solid clusters. The relaxation time is computed through the two-time bond-orientational  correlation function
\begin{equation}
F_4(\vec k,\tau) = \frac{\langle \delta\widehat Q_4(\vec k,t+\tau) \delta \widehat Q_4(\vec k,t)^* \rangle}{N}, \label{defSk4tau}
\end{equation}
where $^*$ stands for the complex conjugate and
$\delta \widehat Q_4(\vec k,t) = \widehat Q_4(\vec k,t) - \langle \widehat Q_4(\vec k,t) \rangle$.
Our results show that for low wavevectors $F_4(\vec k,\tau) \approx F_4(\vec k,0) \exp(-\tau/\tau_4(k))$, from which the relaxation time $\tau_4(k)$ is measured.  Here, we also obtain a critical-like behavior, which is presented in Fig. \ref{fig3}b. The best fit is obtained for 
$ \tau_4/T = \tilde c \varepsilon^{-\nu_{||}}$ with $\nu_{||}=2$, for which the adjusted critical acceleration is $\Gamma_c^{\rm C2} = 5.12\pm0.07$. The relaxation time also seems to saturate for small $\varepsilon$, which occurs at smaller $\varepsilon$ for lower $k$, that is for fluctuations of larger size. The inset of Fig. \ref{fig3}b confirms that $\tau_4 \sim (\xi_4)^z$, with a dynamical exponent $z = \nu_{||}/\nu_{\bot}=2$. 
As usual, there is critical slowing down in the dynamics. As a consequence, close to the critical point, stationary states are obtained after a long relaxation has taken place. Taken that into account, all $\Gamma$ ramps for C2 are slow. Also, averages are taken for long times.

\paragraph{Critical dynamics.} Five critical exponents have been obtained from the analysis of the order parameter. In the standard notation of critical phenomena these are: $\beta=1/2$, $\gamma=1$, $\eta=0.67$, $\nu_{\bot}=1$, and $z=2$. 
In equilibrium, the scaling hypothesis predicts relations among the critical exponents. It is worth mentioning that the relation $\gamma=(2-\eta)\nu_{\bot}$ is not satisfied, while $\alpha+2\beta+\gamma=2$ and  $\nu_{\bot} D=2-\alpha$ ($D=2$ is the spatial dimension) can be satisfied simultaneously if $\alpha=0$. This exponent, associated in equilibrium to the specific heat divergence, has no interpretation out of equilibrium. 

The order parameter in the present case is a non-conserved complex scalar field. Its dynamics, however, is not expected to be autonomous even close to the critical point as density fluctuations are needed to create the ordered phase. Although it has been shown that the transition dynamics is mediated by waves~\cite{clerc2008}, momentum density decays fast due to friction. Therefore, the most appropriate description in the theory of dynamical critical phenomena is model C, in which a non-conserved order parameter is coupled to a conserved non-critical density~\cite{hohenberg}. In this case~\cite{hohenberg, Ccorrect} and in extensions to non-equilibrium dynamics~\cite{noneqC} the dynamical exponent is predicted to be $z=2+\alpha/\nu_{\bot}$, consistent with the measurements if $\alpha=0$.

\paragraph{Conclusions.} We have demonstrated that the non-equilibrium solid-liquid transition that occurs in a shallow, quasi-two-dimensional granular system can be of either first or second order type depending on the vertical height and filling density. 
This seems counterintuitive, because it is widely believed that a solid-liquid phase transition can only be of first order. However, motivated by observations inside carbon nanotubes recent molecular dynamic simulations show that in confined water nanofilms the transition to a solid phase can be either of first or second order, depending on the filling density~\cite{stanley}.
In our experiments, for both cases density fluctuations do not show strong variations at the transition.
On the contrary, local order varies strongly, either abruptly in the first order type transition, or continuously in the second order type configuration. The continuous transition presents critical-like behavior, with exponents consistent with model C of dynamical critical phenomena.

\paragraph{Acknowledgments.} 

We thank M. Cerda, D. Risso, S. Ponce, J. Silva and S. Waitukaitis for valuable technical help and discussions.  This research is supported by Fondecyt Grants No. 1090188 (G.C. \& N.M) and No. 1100100 (R.S.), and Anillo grant ACT 127.


\begin{thebibliography}{99}
\bibitem{resQ6} A similar expression can be written for a 6-fold
bond-orientational parameter.  We have verified that for the current system
and forcing parameters its related global average and Fourier representation
do not vary.
\end{thebibliography}
\end{document}